# A Survey on Deep Packet Inspection for Intrusion Detection Systems

Tamer AbuHmed[1], Abedelaziz Mohaisen[2],[*] and DaeHun Nyang[1]

[1]Information Security Research Laboratory, Inha University, Incheon 402-751, Korea
[2]Electronics and Telecommunication Research Institute, Daejeon 305-700, Korea
tamer@seclab.inha.ac.kr, a.mohaisen@etri.re.kr, nyang@inha.ac.kr

**Abstract**

*Deep packet inspection is widely recognized as a powerful way which is used for intrusion detection systems for inspecting, deterring and deflecting malicious attacks over the network. Fundamentally, almost intrusion detection systems have the ability to search through packets and identify contents that match with known attacks. In this paper, we survey the deep packet inspection implementations techniques, research challenges and algorithms. Finally, we provide a comparison between the different applied systems.*

## 1 Introduction

The enormous attacks from the Internet like viruses, spam, software vulnerabilities and many of attacks spots make protection methods an important way to prevent and save the human efforts from destruction. Therefore, a variety of methods have been used to protect data. These methods began with using cryptography, policies, firewalls,IDS and finally with intrusion prevention systems (IPS) [42]. IDS and IPS are considered as the second defense line against the outsider attack which do not know the cryptographic information. Besides, they work as the first defense line against insider attacks who can bypass the cryptographic system.

The DPI is a core component for many systems plugged in the network including proxies, packet filters, sniffers, IDS, and IPS. Network components use DPI as an essential inspector where it is applied in different layers of the OSI model. Unlike the early beginnings of using DPI where it was applied in only one layer depending on the header (e.g., proxies and firewalls etc.), nowadays, layer-independent attacks force us to inspect attacks in all the layers. According on the intrusion detection literature, efforts to obtain a fast implementation can be categorized into two main categories [31]: (1) design of an efficient data structure with optimized memory access rate, and (2) design of high throughput algorithm to process intruder signature.

In this paper, we survey the deep packet inspection algorithms and their usage in the several existing technologies which are used for intrusion detection systems. The rest of this paper is organized as follows: section 2 introduces an overview on the challenges and goals (or simply objectives) of using the deep packet inspection for efficient intrusion detection systems. Section 3 and section 4 introduce both the software and hardware implementations of DPI systems, respectively. Section 5 overviews the finite state machine, section 6 introduces a comparison between the existing technologies and architectures, and finally section 7 draws concluding remarks.

## 2 Challenges and Goals

The design and implementation of the deep packet inspection has several challenges which harden the its advancement process. Also, there are several ultimate goals and design objectives that are always considered when we make a new DPI design. In this section, we list the different challenges and design objects.

### 2.1 Deep Packet Inspection Challenges

When the DPI becomes mean to detect the intrusion, there are several challenges related to applying it on the network. In the following, we summarize these challenges.

1. **The search algorithm complexity:** the complexity of the algorithm and the operations of

comparison against the signatures of intruder decrease the throughput of the system. Thus, search algorithms are the main focus point in DPI researches, whereas matching process is resource consuming. For example, the string matching routines in SNORT [35] account for up to 70% of total execution time and 80% of instructions executed on real traces[4].

2. **Increasing number of intruder signature:** according to the verity of attacks, the needs for new intruder signature increase. Therefore, the large number of signatures makes the task of IDS harder whereas the matching process must inspect traffic against all attacks fingerprints.

3. **The overlapping of signatures:** the signatures of attacks usually are not general so the signatures can be categorized into groups according to common properties like protocol type. For example http packet in snort [35] has 1096 signatures. Therefore, there is a need for process the packets before matching process.

4. **The Location of signature unknown:** due to verity types of attacks on different types of applications, the pattern of intruders is not localized in specific place in the packet which means that the IDS must inspect all the payload of the packet against the attacker signatures.

5. **Encrypted Data:** the data which is encrypted cannot be inspected by DPI. However, there are some solutions to overcome this problem by plugging the DPI component behind the decryption device.

The DPI system as we mentioned before has many challenges and in the same time it have to provide the requirements for network need. There are two main requirements that should be satisfied on DPI system, more detail will be provided in subsection 2.2,which is:(1) the high speed of processing the packets which affects the throughput of the system and manages the core speed of the network (10 Gbps-40 Gbps) and the edges speed (1 Gbps). (2) The low cost for DPI system as memory, and power consumptions.

## 2.2 DPI Design Objectives

DPI systems have to satisfied specific objectives to sustain the traffic rate and intrusion signatures growth. Hence, we conclude some objectives which have to satisfy in DPI architecture as following [45] [40]:

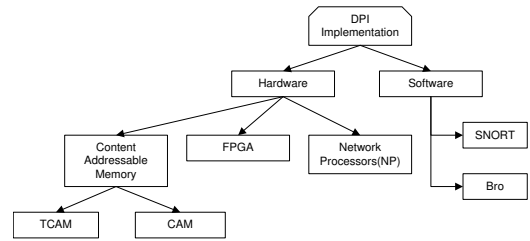

**Figure 1. DPI implementations**

1. **Deterministic performance**: the architecture has to operate and process traffic stream independently of signature characteristics or traffic characteristics. So, the system has to manage traffic in worst case in software and hardware based systems.

2. **Memory efficiency**: memory access time is one of the main bottlenecks in DPI system in software implementations meanwhile, it is critical in hardware design as access time and memory scarcity. Thus, high memory efficient design is preferable.

3. **Dynamic update**: this objective is very important in hardware based design to add and remove intruder signature to system without affect system operation.

4. **Signatures**: DPI system support fixed intruder patterns and regular expression. Also, the system can deal with all types of intruder patterns [20] which we will illustrate in the literature in section 4.4.

5. **Scalability**: scalability is not big issue in software based system. On the other hand, it is critical in hardware based systems. Thus, hardware design has to support unlimited number of signatures.

6. **Additional functions**: DPI system can support another function like; multi traffic's sessions inspected separately, not only inspect the intruders but also allocate it, and customize signatures subsets or entire signature to inspect.

## 3 Software Deep packet Inspection systems

There are many packet scanning applications that require deep packet inspections. Here, we review three popular ones: SNORT [35], Bro [10] and Linux L7-filter [28]. SNORT and Bro are two popular intrusion detections systems, while L7-filter is an application for application layer protocols analysis which makes packet classification based on application layer data. These systems are all open source systems, which allow us to perform a detailed analysis and show their abilities and constraints.

## 3.1 SNORT Intrusion Detection System

SNORT is an open source intrusion detection system which used for protocol analysis and full packet inspection against intruder signature. The SNORT system processes the traffic of packets on multi stages as illustrated in Figure 2 [47]. SNORT system and all common IDS use method called analyze-normalized-matching (ANM) [32]. SNORT use many string matching algorithms, on of them is Boyer Moore (BM) algorithm which we will talk about it in literature about matching algorithms in section 4.1. SNORT rule may contain header and content fields where the header part checks the protocol, source and destination IP address and port, and the content part scans packets payload for one or more patterns. Rules with more than one pattern are called correlated rules. Furthermore, rules can also contain negation patterns, which mean negation of patterns stands for no occurrence of the pattern. The matching pattern may be in ASCII, HEX or mixed format. HEX parts are included between vertical bar symbols "j" as an example of a Snort rule is [35]:

```
alert tcp any any -> 198.165.200.24/32 111
(content: "idcj|3a3b|j"; msg: "mountd access";)
```

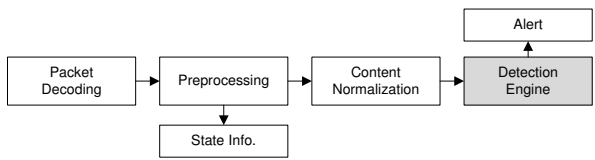

**Figure 2. SNORT Process Stages**

## 4 Hardware Implementation

As a need to speed up the inspection process, the hardware (HW) implementations always appear as a preferable solution for high speed DPI implementation. However, the different requirements for DPI provide limitations to perform the deep packet inspection in HW. The limitation refers to the large number of signature, complexity and overlapping of signatures and finally the high rate of signature update and addition. Therefore, the HW solution has to satisfy the previous requirements by special properties which are as follows:

1. Use of high degree of pipelining to support inspection for large number of intruder patterns.

2. The HW component must have high degree of processing capability to manage complex patterns with LAN speed (e.g., 10 Gbps).

3. It must be configurable HW to be suitable for changing situation of intruder patterns.

4. It must be design to be capable of update or add a new pattern without turning off the DPI component.

The hardware implementation can be categorized into three depending on the used technologies in that implementation as follows:

1. Ternary content addressable memory (TCAM) implementation [41]

2. Field-programmable gate array (FPGA) implementation [17]

3. Multi-core processors [22]

However, each implementation has its advantages and limitations which as we will see later when we detail each implementation. In general, multi-core processors implementations are considered the best preferable among the implementations due to its programming flexibility. On the other hand, the TCAM is preferable when the speed is considered.

## 4.1 Matching Algorithms

The matching for pattern depends on the algorithmic way to process the data and return the result of existence of the pattern or not in considerable time. Accordingly, many algorithms have been introduced to perform string matching. Though, the string matching algorithms always suffer from two factors that affect the throughput of processed data. The first factor is the computation operations to make comparison between the pattern and the data and second is the number of patterns that need to be compared with the traffic of the incoming data. Historically, the first string matching algorithm was the brute force (BF) algorithm which compares the first character in the pattern with the data stream. If the a single charter match, BF compares it with the next character of the pattern and so on. Finally, if the whole pattern is finished, it issues the pattern matching results.

Later on, many algorithms appear to increase the performance of matching. These algorithms can be categorized according to the implementation as software based, HW based or mixture of both implementations. Briefly, there are a lot of algorithms for pattern matching. However, the most famous software based algorithms are Knuth-Morris-Pratt (KMP) [24], Boyer-Moore (BM) [9], Aho-Corasick (AC) [1], AC_BM algorith [14], Wu-Manber [48], and Commentz Walter (CW) [15]. We will summarize the concept behind selected algorithms and their implementation, design, and applicability for DPI. On the other hand, most known HW based algorithms are the parallel Bloom Filters [17], CAM (content addressable memory), TCAM, and finally FPGA implementations.

**KMP Algorithm:** the Knuth-Morris-Pratt (KMP) algorithm [24] came as an enhancement for the brute force algorithm which was we introduced before as the early work for pattern matching. The improvement of KMP over the BF is performed by skipping characters when the mismatch occurs in the comparison phase. This skipping for characters depends on preprocessing phase of KMP to the patterns. The result of the KMP is somehow similar to the finite automata for patterns representation in which depending on every match and mismatch a certain jump over the input stream occurs. Additionally, KMP [24] and BM [9] algorithms are designed for single pattern searching.

If the pattern length is $m$ bytes, the complexity of the matching algorithm will be of $O(m + n)$ matching this pattern in an $n$ bytes stream. If there are $k$ patterns, the search time will be $O(k(m + n))$ according to that the single search is performed $k$ times. In [7], Baker and Prasanna implemented a hardware based DPI architecture for KMP algorithm to exploit the HW parallelism and reduce the complexity of the above bound.

## 4.2 Bloom Filter

The Bloom filter is a technique to generate a structure that compresses the pattern string as $s$ hashed value. After that, the same hash function that produced the patterns is used to make the dependences from the input traffic. This method has been applied firstly in intrusion detection system by Dharamapurikar et al. [17] and his implementation was on FPGA. The system implementation achieves a throughput of 2.12Gbps. Bloom filters are very elegant in representing set membership, but have two potential drawbacks. First, they require multiple hash functions and memories, and second, they give an approximate match answer since they allow false positives.

## 4.3 Content Addressable Memory

Nowadays, the most popular HW techniques which are used in commercial packet inspection products are content addressable memory (CAM) [41]. The CAM is a special memory that makes parallel comparison for its contents against the input value and returns the address of match entry. Hence, the CAM is considerably fast and has many demanded properties such as high access speed near 4 nano-second, the search time complexity is $O(1)$ and bounded by a single memory access. However, CAM does not make longest prefix matching which is essential for many DPI patterns that have the same prefix. Therefore, it is suitable for deterministic fixed-length matching.

Also, because of the above shortage of CAM, a new HW component was developed by the name of Ternary CAM (or simply, TCAM). TCAM memory stores the data with three logical values (*i.e.,* 0, 1, ? don't care) and its circuit diagram construct as illustrated in Figure 3(b) [41]. Furthermore, each entry stores the value which is considered to be intruder signature and entries arranged in descending index as illustrated in Figure 3(a) [41].

As a result of the previous properties, for CAM and additionally to Longest-Prefix Matching, TCAM became as backbone for many network devices that depend on packet inspection. For example routers and switches primarily use TCAMs to perform forwarding lookups for Internet Protocol addresses. TCAMs can be also used in devices that support packet classification, network address translation, route lookups in storage networks, layer 4 to layer 7 switching, server load balancing, label switching, high performance firewall functions and finally in network intrusion detection system (NIDS) and network prevention system (NIPS) that depend on DPI techniques.

However, TCAM has some general disadvantages which are as following [41]:

1. High cost per bit relative to other memory technologies, it's about 30 times SRAM per bit.

2. Storage inefficiency.

3. High power consumption. It is about 180 times than SRAM per bit and the power consumption proportional with number of entities which has been searched on memory lookup.

4. Limited scalability to long input keys.

The special disadvantages for DPI are as follows [29]:

1. Range Representation Problem: TCAM can represent prefix of patterns in easy way (*e.g.* "atta XX"

catch any word start with atta and two letter after) but rang signature which catch sub-word and after arbitrary number of character catch the reminder sub-word consumes more entries in TCAM.

2. Multi-match Classification Problem: Return back all the matching results of all matching entries of TCAM, not just the highest priority entry of TCAM.

**Bitwise CAM:** In [50], CAM hardware has been implemented based on a tree-based content addressable memory structure called "Bitwise CAM", which involves HW sharing at bit level in order to exploit powerful logic optimizations for multiple strings represented as a Boolean expression. The design can run at a rate of approximately 2.5 Gbps per second, and is approximately 30% smaller in area when compared with published results. Also, authors functionalized the parallelism in the design of an extended system.

## 4.4 TCAM implementations

In literature of TCAM's contribution in DPI, Yu et al. [20] have been the first to design scheme that deals with all types of intruder patterns which we will discuss later. In [20], they implement a scheme for IDS that handles the intruder's signatures with deeply analysis to intruder's patterns. The scheme categorizes intruder patterns into two types: complex patterns such as long patterns, patterns with negation (which means no existence of specific patterns on traffic) and correlated patterns (which means patterns separated with specific number of arbitrary characters). Additionally, there are another type which is a simple pattern.

The work by Yu et al. discusses scheme and algorithms to deal with each type of pattern and how to plug it into TCAM. The scheme uses SRAM memory as partial hit list (PHL), which consider slow in access comparing to TCAM, to store detection of partial correlated patterns encounter in traffic. Nonetheless, the scheme has bottleneck when the intruder intentionally send packet that make PHL access rate very high and then effect the system throughput. That is due the need of multi memory look up.

According to the simulation, this scheme can be operated on 2 Gbps traffic. The implementation of Yu et al. in [20] suggests lookup on TCAM entries for each new character. Thus, the input of n character requires the complexity of $O(n)$ lookup over TCAM. On the other hand, Jung et al. in [38] presented a scheme in which jump are made over the input traffic by window slide size $m$ which is called jumping window scheme and match the intruder signature over single packet.

It reduced the number of TCAM lookup over $n$ input character to $O(n/m)$ and provided throughput of 10 Gbps using 2,394 SNORT rules. Also, Sung et al. in [39] extended the jumping window scheme to work over multi packets intruder signatures.

## 4.5 Multi-core Processors Implementations

Multi-core processors' implementations are preferable for designing IDS due to flexibility. However, multi-core processors still have limitation in number of processors and size of on-chip memory which affect efficiency of IDS implementations on it. In the following, we will introduce a survey on a part of the efforts been performed to implement IDS on network processors (NP) which is a type of multi-core processor implementation.

In [16], Bruijn et al. developed the SafeCard desgin which is a framework for network-based intrusion prevention at the network edge which is able to cope with all levels of abstraction and can be easily extended with new techniques. Furthermore, it is capable of reconstructing and scanning TCP streams at Gbps rates while preventing polymorphic buffer-overflow attacks.

Additionally, the CardGuard by Bos et al. in [8] uses IXP1200 network processor as IDS and achieved few hundred Mbps Ethernet performances when scanning payloads of TCP connection. In [34], Singh et al. introduce Early-bird prototype which consists of sensor to detect attacks and aggregator for administrative reporting and control. Early-bird can cope with 200 Mbps without packet dropping.

In [12], new work has been introduced by Chris et al. as a combination between IXP network processors and Xilinx Virtex FPGAs to build IDS.

## 5 Finite State Machine

One of the most important tools for the design of hardware implementation for the DPI is the finite state machine (FSM). The FSM implementation is classified into two categories which are the deterministic finite automata (DFA) and nondeterministic finite automata (NFA). In this section, we introduce a survey of the research that has been performed on the FSM including the two categories.

### 5.1 Nondeterministic Finite Automata

Nondeterministic Finite Automata (NFA) is a directed graph which has nodes called states and labeled edges to connect the states. More specifically, the NFA has initial state and one or more final states. Moreover, the edges can be labeled with single characters or null ($\phi$) which mean that multiple states can be active simultaneously in an NFA. The NFA is very useful in parallel processing because it can process input character in multi branches of NFA and may output multi acceptance state for input on the contrary of DFA [21].

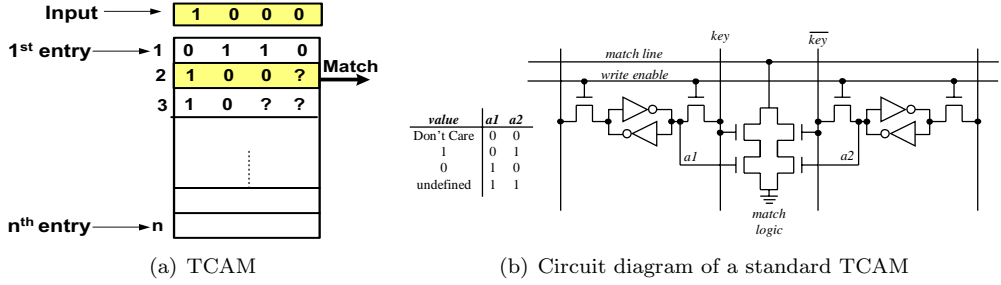

**Figure 3. (a) TCAM (b) TCAM cell value(0,1,?) encoded by two register a1,a2**

For its usability, there are many efforts to construct DPI systems which depend on NFA. In [33], Reetinder et al. were the first how to use the NFA to construct regular expressions in given text using FPGAs. To match a regular expression of length $n$, a serial machine requires $O(2^n)$ memory and takes the time complexity of $O(1)$ per text character. However, they proposed an approach that requires the $O(n^2)$ space and still process a text character in $O(1)$ time (one clock cycle). Additionally, they presented a simple and fast algorithm that quickly constructs the NFA for the given regular expression. Fast NFA construction is crucial because the NFA structure depends on the regular expression, which is known only at runtime. Furthermore, in [13], Clark et al. implemented FPGA based multi character decoder for DPI which based on NFA.

### 5.2 Deterministic Finite Automata

The Deterministic Finite Automata (DFA) consists of a finite set of input symbols (which are denoted as $\sum$), a finite set of states, and a transition function to move from one state to the other denoted as $\partial$. In contrast of NFA, DFA has only one active state at any given time [21].

**Regular Expression:** The regular expression is required as a need for packet payload inspection to different protocols packets. It introduces a limited DPI system to deal with all packets structures. As the result of this limitation, state-of-art systems have been introduced to replace the string sets of intrusion signature with more expressiveness regular expression (regexp) systems. Therefore, there are several content inspection engine which have partially or fully migrated to regexps including the those in Snort [35], Bro [10], 3com's TippingPoint X506 [42], SafeXcel [19], and Cisco systems' [23]. However, using the regexp to represent patterns includes converting this regexp to Deterministic Finite Automata (DFA) [21]. This DFA is represented in the DPI systems as table. This table represents the states and transitions of DFA as records which mean that the expansion of memory table of DFA of regexp depends on the size of DFA.

Experimentally, DFA of regexp that contains hundreds of pattern yields to tens of thousands of states which mean memory consumptions in hundreds of megabytes. As a solution of one of the common problems of HW based DPI solutions is the memory access because the memory accesses for the contents of the off chip memory are proportional with the number of bytes in the packet.

In [26], Kumar et al. noted that the implementation for the regexps of intruder signatures consumes much memory and there should be a way that reduces the regexp memory consumption without increasing the number of memory lookup to operate DPI system which is considered an additional problem due to the related lookup delay. To reduce the memory access, they also introduced a delayed input DFA D$^2$FA which tries to compact the traditional DFA for regexp according to that they note some states in DFA that had the same outgoing transition. For example, if there are two states $s_1 1$, $s_2$ that introduce transition to the same outgoing set of stats ($S$) for set of input characters C, this transition can be eliminated from state $s_1$ by default transition DT to $s_2$.

According to this assumption, the state $s_1$ can maintain all the transition of state $s_2$ via state $s_1$ and then passing to next state. D$^2$FA constructs a compact DFA which decreases the memory consumption by DFA. However, compacting the memory representation by default transition leads to manipulation of multiple default transition before going to the next state. Manipulating multiple DTs means that multiple memory accesses are required which decrease the DPI process throughput. However, the they (i.e., Kumar et al.) found that applying $D^2FA$ can reduce the memory usage dramatically about 95% which helps to implement DPI in an On-chip memory and that leads to high bandwidth in memory access and decreases the effect of multi-transition access by DTs to process input character. The construction of $D^2FA$ from DFA is NP-hard. Therefore, they introduce heuristic algorithms to find $D^2FA$ with balancing between the depth of DTs and the memory consumption for $D^2FA$. $D^2FA$ construction heuristic based upon maximum weight spanning tree creates long default paths [25].

**Table 1. Comparison between Existing Architectures**

| Algorithm / Component | Implementation Device | Throughput (Gbps) |
|---|---|---|
| Parallel Bloom Filters [17] | FPGA XCV2000E | 2.46 |
| Aho-Corasick [3] | FPGA | 12.35 |
| TCAM [20] | TCAM | 2 |
| Aho-Corasick [44] | – | 8 |
| TCAM/FPGA [43] | Xilinx Virtex2 | 10 |
| nnnnn/SRAM [2] | – | 14 |
| Selective multi-character transitions /FPGA [37] | Xilinx XC2V6000-6 | 14 |
| B-FSM/(FPGA or ASIC) [45] | Xilinx Virtex-4 | 10$\sim$20 |
| nnn/SRAM [3] | FPGA/ASIC | 1$\sim$20 |
| RTCAM [46] | TCAM | 12.35 |
| Pre-Decoded CAM [36] | Virtex 2-6000 | 9.7 |
| Quad Bloom Filter/FPGA [6] | Xilinx Virtex4 | 20.4 |
| BITWISE CAM [50] | FPGA Xilinx XC2V8000 | 2.5 |
| FPGA [18] | Virtex-4 | 10 |
| UCLA Packet/FPGA [11] | Xilinx Spartan 3-XC3S2000 | 3.2 |
| NFA/(FPGA and IXP) [12] | Xilinx Virtex2-6000&IXP 2400 | 1 |
| GaTech Decoder Trees/FPGA [13] | Virtex 2-8000 | 2 |
| WashU Bloom/FPGA [5] | Virtex 4-100 | 20.4 |
| Hash Function [49] | Xilinx Vertex-II Pro XC2VP70 | 2 |
| Hash Function and CRC [30] | Xilinx Vertex2 | 2.712 $\sim$ 4.560 |
| TCAM/Network Processor [38] | Network Processor IXDP28xx [22] | 10 |

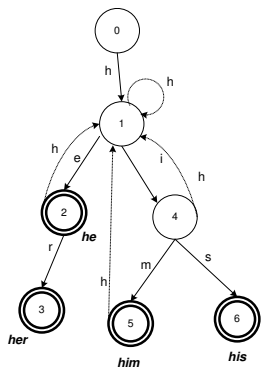

**Figure 5. Aho-Corasick DFA for patterns "he", "she", "his", and "her", we did not include all failure edges for simplicity.**

In [27], which is also by Kumar et al., a new representing for regexp has been developed as an alternative to $D^2FA$ which has the property of being compressed from $D^2FA$ and improve the ability of processing multi DTs to handle input characters by introducing more information in state identifiers. Content-addressed $D^2FA$ $CD^2FA$ replaced state identifiers with content labels that include part of information that would normally be stored in table entry for the state. The main idea of $CD^2FA$ is exploit the $D^2FA$ compaction to DFA but on the other hand is to overcome the multi TDs traversing to manipulate the input. Notwithstanding, $CD^2FA$ need to increase the size of the states label to hold more information about the next state and DTs. So that, there are two objectives to satisfied: First, to ensure that states have few labeled transitions. Second, to ensure that default paths are as small as possible.

According to experimental evaluation, $CD^2FA$ go beyond uncompressed DFA. Furthermore, $CD^2FA$ with 1KB cache achieves double throughput than uncompressed DFA and with 10% of memory requirement.

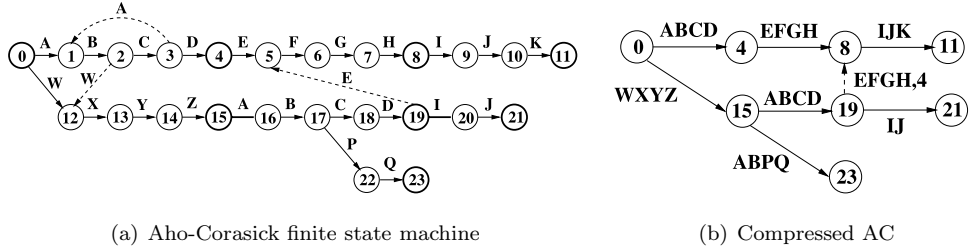

(a) Aho-Corasick finite state machine    (b) Compressed AC

**Figure 4. Compressed AC for high speed DPI**

**Aho- Corasick Algorithm:** Aho- Corasick Algorithm (AC) [1] is one of the well known algorithms for multi-string (patterns) matching by encoding intruder patterns in FSM in a preprocessing phase. After that, the generated FSM has root state which represent that no string have been matched or even partially matched and all patterns characters enumerated from root. If any pattern has same prefix, it means that the pattern shares a common prefix also with the corresponding set of parent nodes in the tier. Figure 5 shows a example of the AC FSM construction for patterns "he", "she", "his", and "her". However, AC construction is memory consumption as a result of the huge number of failed transitions that proportional with the number of patterns in FSM. Thus, classical AC takes more storage than it is likely to fit in a on-chip SRAM or the cache of a processor [44].

Additionally, In [3], Mansoor et al. constructed a compressed finite state machine that encodes all the intrusion patterns and makes state transitions on multiple (at most $k$) input characters. Therefore, they start constructing Aho-Corasick DFA as in Figure 4(a), then they create an equivalent state machine called the compressed DFA as illustrated in Figure 4(b) where it has transitions on multiple input characters by combining $k$ consecutive states of Aho-Corasick DFA. Conversely, in [40], Lin et al. proposed a new construction for AC by splitting the input character to bits and constructing small blocks that represent portion of rules with portion of bits for each rule. This construction exploits a speedy on-chip memory to upload the small block of the system and speed up the overall system throughput.

## 6 Comparison between Existing Modules and Implementations

In this section, we introduce a comparison between recent applied IDS with different hardware implementations. Our comparison focuses on the algorithm, type of hardware implementations which are used in designing the DPI architecture and the resulting throughput as illustrated in Table 1. However, other related properties including the required memory and other specifications might be referred in the corresponding reference.

## 7 Conclusion

In this paper, we introduced a survey on some of the existing and on-going research works on DPI. Our survey included the challenges and ultimate goals behind the design of the the DPI and its implementations. Also, we introduced an overview of the existing implementations including both the software and hardware. As the finite state machine (or automata) is an important component of the hardware design, we considered the its different classified types and the ongoing research being performed on each type. Finally, we introduced a concluding comparison between the existing modules and hardware implementations and relating this comparison to the achieved throughput.

We believe that this area of research is still active and several works need to be performed on the different sides of the implementation (hardware and software) in addition to the design of fast matching algorithms that fit to the increasing demanded throughputs. Our survey is the first step for putting the readers into the the DPI systems and the open research topics in the field.